\documentclass[a4paper,12pt]{article}
\usepackage[dvips]{graphicx}
\input epsf.tex

\begin{document}

\title{Production of Super Heavy Elements at GANIL: \\
present status and perspectives}

\author{
S.~Gr\'evy
$^{a}$ and FULIS
   collaboration:\\
N.~Alamanos$^{b}$, N.~Amar$^{a}$, J.C.~Ang\'elique$^{a}$, R.~Anne$^{c}$,\\
G.~Auger$^{c}$, F.~Becker$^{c}$, R.~Dayras$^{b}$, A.~Drouart$^{b}$,\\
J.M.~Fontbonne$^{a}$, A.~Gillibert$^{b}$, D.~Guerreau$^{c}$,
F.~Hannape$^{a}$,\\
R.~Hue$^{c}$, A.S.~Lalleman$^{c}$, T.~Legou$^{a}$, R.~Lichtenth\"aler$^{d}$,\\
E.~Li\'enard$^{a}$, W.~Mittig$^{b}$, F.~De~Oliveira$^{b}$, N.~Orr$^{a}$,\\
G.~Politi$^{e}$, Z.~Sosin$^{f}$, M.G.~Saint-Laurent$^{c}$,\\
J.C.~Steckmeyer$^{a}$, C.~Stodel$^{c}$, J.~Tillier$^{a}$,
R.~de~Tourreil$^{c}$,\\
A.C.C.~Villari$^{c}$, J.P.~Wieleczko$^{c}$ and A.~Wieloch$^{f}$.\\
\\
{\small $^{a}$LPC, IN2P3-ISMRa-Universit\'e, Caen, F-14050, France}\\
{\small $^{b}$DAPNIA/SPhN, CEA Saclay, Gif-sur-Yvette, France} \\
{\small $^{c}$GANIL, Caen, France}\\
{\small $^{d}$IFUSP, S\~ao Paulo, Brazil}\\
{\small $^{e}$Universit\`a di Catania, Italy}\\
{\small $^{f}$Ins. Fyziki Uniw., Krakow, Poland}\\
\\
{\small proceeding to the ASR2001 Conference, to be published}\\
{\small in the Journal of Nuclear and Radiochemical Sciences (june 2002)}\\
 }
\date{}

\normalsize

\maketitle

\abstract{ Experiments on the production and study of superheavy nuclei have
been undertaken at GANIL taking advantages of the powerful velocity filter
LISE3 and the high intensity ECR ion sources. A complete set-up has been built:
- reaction chamber containing large rotating wheels, - slits and beam profiler
at mid-filter, - detection chamber with time-of-flight and Silicon detectors
for identifying evaporation residues and their decay products. The response was
checked via fusion reactions with known cross sections producing known
$\alpha$-decay chains and fission fragments. Excellent transmission of
evaporation residues and rejection factor for the primary beam were obtained.
This device is presented together with the results obtained in our last
experiments and our future plans are discussed.
}%

\maketitle

\section{Introduction}

An experimental program dedicated to the study of Super Heavy Elements (SHE)
started at GANIL in order to take advantages of the performances of the
accelerator and of the powerful velocity filter LISE3.

Indeed because of the very low cross sections for the production of SHE (from
1~$\mu$barn for Z=100 to 1 pbarn for Z=112)~\cite{ref1}, very high intensity
ECR ion sources delivering about one particle$-\mu$A (6.27x10$^{12}$
projectiles/sec) or more should be used. The second important point is the
separation between the projectiles and the complete fusion evaporation residues
(ER) emitted at forward angle after the production target. Originally, the
LISE3 Wien filter~\cite{ref2} was designed to improve the separation of exotic
nuclei produced by fragmentation of the projectile and selected by the LISE
spectrometer. Then, an important work has been done to adapt it to fusion
experiments improving its ability to work with very high primary beam
intensities.

An experimental setup composed of a dedicated rotating target, beam
diagnostics, time detectors and a set of Si-detectors (for the implantation and
decay products) has been developed. Finally a digital electronics is under
development to reach extremely short decay times in the order of few
$\mu-$seconds.

The response of the whole setup was checked via fusion reactions with known
cross sections and $\alpha$ decay chains: $^{204,206}$Fr$_{87}$ nuclei formed
via $^{86}$Kr+$^{nat}$Sb with cross sections of 10 to 300
$\mu$barn~\cite{ref4}. These equipments were then used for an experiment
dedicated to the system $^{86}$Kr on $^{208}$Pb and the transmission and the
rejection factor of the Wien filter were precisely measured in an experiment
dedicated to the production of the Sg element (Z=106) through the system
$^{54}$Cr+$^{208}$Pb. Comparison with known values for the excitation energy
function~\cite{ref1} allowed to check the excellent transmission together with
an high rejection factor.

An experimental program has been established for the production of new elements
and/or new isotopes with two principal axes: $\it -i-$ the use of the inverse
kinematics with a beam of Lead and $\it -ii-$ the use of the neutron-rich
$^{76}$Ge as projectile. Another program dedicated to the study of the
properties of SHE has also started recently.

\section{Experimental setup}

A general overview of the experimental setup, from the reaction chamber to the
detection setup is presented in Figure~1.

\subsection{Reaction chamber}

The reaction chamber contains two 67~cm diameter wheels bearing 35 targets
(5.0*1.5 cm$^2$ each) and rotating up to 2000 RPM, allowing targets with low
melting temperature (like Pb or Bi) to sustain very intense primary beams. The
second wheel support carbon stripper foils in order to re-equilibrate the
charge state of evaporation residues. The time structure of the beam is
synchronized with the rotation. Two smaller wheels allow us to run with targets
having higher melting temperatures. A Si detector continuously monitors the
status of each target via elastic scattering reactions.

\subsection{Wien filter}

The LISE3 Wien filter has crossed magnetic and electric fields in a rather
compact geometry and its angular acceptance is $\pm$36$\pi$mrad. It is divided
into two identical halves and is followed by a dipole magnet. Several
modifications were found to be necessary for fusion experiments: -$\it{i}$- the
upper plate in the first half of the filter was moved up by 2~cm in order to
deflect the beam without hitting the electrode. The beam is now stopped on a
new water-cooled collimator, -$\it{ii}$- two pairs of independently movable
slits and a beam profiler were installed at mid-filter. An additional
background suppression can be performed using the last dipole magnet located
between the Wien filter and the detection chamber. The transmission of fusion
nuclei from the target position through the quadrupoles, filter and final
dipole was studied via a simulation code~\cite{ref3}. The optics are checked
and the Wien filter is calibrated using low energy ions (0.35~MeV/u to have the
same speed than ER's after the target).

\subsection{Detectors}

The detection setup (see Figure 2) is composed of two Galottes (carbon foils +
MCP's time detectors) and 13 Si detectors. The time detectors are used in order
to tag all the events (ER's or background) coming from the beam. Since the
efficiency of Galotte for light particles might not be 100\%, two of them are
used. Then, the "implantation events" are those with an implantation energy
signal in coincidence with both Galottes whereas the "$\alpha$" or fission
events are those in anti-coincidence. The implantation detector is a 300$\mu$m
thick Si 48x48 strips. The distance between the strips is reduced
($\sim$45$\mu$m) to have a dead zone inferior to 5\%. Two of them are mounted
on a movable arm. When emission of $\alpha$-particles or fission fragments is
identified via electronics, the beam is immediately stopped for a short while
during which the first implantation detector is moved out of the beam and the
second implantation detector comes in, then the beam is sent again. This
technique allows us to wait for very long decay times in the first implantation
detector, being completely free of background coming from the beam. The ER's
are implanted at a depth of approximately 10 $\mu$m which is smaller than the
range in Si for $\alpha$ particles with energy above 2.5 MeV. Then, in order to
measure the full energy of the $\alpha$ particles emitted at backward angles
which escape the implantation detector, a set of 8 Si detectors is mounted in a
'tunnel' geometry. With this configuration, the efficiency for full $\alpha$
energy goes up from 55 to 95\%. A 'face' Si detector is placed in front of the
'out of beam' implantation detector in order to play the same role, the
efficiency being then 100\%. A veto Si detector is placed after the
implantation detector to eliminate the background of light particles coming
from the beam. All the Si detectors are mounted on copper supports and can be
cooled to reduce the electronic noise.

\subsection{Data acquisition and Electronics}

Specific electronics and data acquisition systems were developed. We use for
the data acquisition a complete VXI system for which the standard dead-time is
around 150$\mu$s. In order to reduce it, a multi-trigger logic with 2 sets of
converters (ADC's) is working in the following way: the hard trigger provided
by the electronics enters into two GMT (Ganil Master Trigger) modules, T1 and
T2. In case of usual counting rate, all the triggers are treated as 'T1' and
the events are registered into the first set of converters. This way is exactly
the same as the usual one used in many experiments. The same signals are also
duplicated into the second set of converters but are not yet considered by the
acquisition. In case of a new trigger comes during the dead time of GMT1, it is
now treated by GMT2 (as a trigger 'T2') while GMT1 continues to deal with the
preceding event. The minimum time between two events is then about 30$\mu$s.
This procedure was tested with a strong $\alpha$-source before the experiment.
The used converters are new 14-bits ADC's which allow to have a good resolution
for $\alpha$-particles (less than 50 keV) together with a large range of
energy, between 300 keV and 250 MeV (necessary for spontaneous fission events).

A fast analysis program allows us to identify 'on line' $\alpha$-decay chains
or spontaneous fission and then to stop the beam and move out the implantation
detector (see sect. 2.3)

A complete read out digital electronics is actually under development to be
able to reach extremely short decay times, in the order of few $\mu$sec keeping
good energy resolution (less than 50 keV) on a large energy range from 1~MeV to
1~GeV. This new digital electronics will be placed after the analog front-end
electronics (preamplifiers) and the main idea of such an approach is to adapt a
dedicated processing to each event on a very large dynamical range and without
limitation due to the recovery time of the preamplifier.

\section{Experimental results}

\subsection{Experiment on $^{86}$Kr + $^{208}$Pb system}

In July 1999, the BGS-Berkeley group reported the observation of 3 events of
the new element Z=118 identified by $\alpha$-decay with chains of
6~$\alpha$'s~\cite{ref5}. The reported cross section was much larger than
expected ($\sim$2~pb) and the same system studied with the velocity filter SHIP
at Darmstadt did not lead to the observation of this element~\cite{ref6},
justifying to repeat this experiment at GANIL in december 99. Since GANIL is
able to deliver an high intensity beam of $^{86}$Kr (15$\mu$A with charge 10+,
i.e. nearly 10$^{13}$projectiles/sec), our idea was to confirm (or not)
Berkeley's result and, in the case of a positive answer, to obtain additional
information: -$\it{i}$- be able to reach shorter decay times (by the use of
fast electronics, see sect. 2.4.) for possible $\alpha$ decay before the
minimum time of 120 $\mu$sec in the Berkeley experiment, -$\it{ii}$- wait for
very long decay times (by the use of movable implantation detectors, see sect.
2.3.), -$\it{iii}$- since ER's may be created in an isomeric state which decays
via electron capture followed by an electron cascade with an unknown half-live
modifying the ionic charge and strongly reducing the transmission in the case
of the GSI experiment, the carbon stripper foils were in our experiment located
at a distance from the target 3 times larger than in the SHIP experiment.

With a total accumulated dose of 1.1x10$^{18}$ ions at 5.27 MeV/u on targets of
300~$\mu$g/cm$^2$, we didn't observe any event corresponding to the decay of
element 118.

The result of Berkeley was recently retracted~\cite{refBer}.

\subsection{Experiment on $^{54}$Cr + $^{208}$Pb system}

In order to measure precisely the transmission and the rejection factor of the
LISE3 Wien filter for SHE's, an experiment dedicated to the production of the
Sg$_{106}$ element through the system $^{54}$Cr on $^{208}$Pb was performed at
GANIL in december 2000. With a primary beam intensity of 40 nAp and 2 incident
energies (4.698 and 4.756 MeV/u), we observed 10 events of $^{261}$Sg$_{106}$
(1 neutron evaporation channel) and 2 events of $^{260}$Sg$_{106}$ (2 neutrons
evaporation channel). The decay chains were identified by position correlation
between the implantation of the ER and the decay events in the implantation
detector. Comparison with the known excitation energy functions~\cite{ref1}
makes possible to deduce a transmission efficiency above 60$\%$ together with a
power of suppression of the primary beam and scattered projectiles of
$\sim$2x10$^{10}$, corresponding to a counting rate of 5-10 Hz in the
implantation detector.

\section{Short and Middle range Perspectives}

\subsection{Inverse kinematics}

\subsubsection{Method}

The use of the inverse kinematics (heavy projectile on light target) has been
tried at GANIL in June 1999 and will be used in spring 2002 on the system
$^{208}$Pb+$^{54}$Cr to produce $^{261,260}$Sg$_{106}$ isotopes~\cite{ref8}.
This technique presents several assets in comparison with direct kinematics

-$\it{i}$- the target thickness is no longer limited by multiple scattering of
the ER's and it becomes possible to use thicker targets covering an excitation
energy range of typically 10 MeV instead of 3-4 MeV. The counting rate is then
maximized and the gain in time can be very important. Indeed, in the search for
new isotopes or new elements, the incident energy depends strongly on the mass
defects which differ in different mass tables by several MeV whereas the
excitation function is rather narrow.

-$\it{ii}$- the ER's are more strongly focussed at forward angles resulting in
a better transmission through the LISE3 filter.

-$\it{iii}$- because of the higher energy after the target, the ionic charge
distribution of ER's is better estimated and its relative width is smaller.
Here again the transmission is favoured.

-$\it{iv}$- finally, the quality of the data is better: due to the deeper
implantation depth, the energy deposited by the ER's is higher, both fission
fragments are stopped and escaping $\alpha$ particles lose more than 2 MeV and
then cannot be missed.

The main drawback of this approach concern the velocity difference between the
beam and the ER's which is much smaller than in direct kinematics. Then, higher
electric and magnetic fields are necessary in the Wien filter. With the new
power supplies installed this year, the LISE3 filter will be able to get the
same spatial separation at mid-filter as in direct kinematics.

\subsubsection{Use of $^{208}$Pb beam}

Several systems can be studied for different purposes (summarized in Table 1):

$\it{-i- New~paths~to~known~isotopes~with~^{208}Pb:}$ Produce the same isotopes
of the odd Z compound nuclei than ER's formed at GSI with a $^{209}$Bi
target~\cite{ref1}. Here, two points are of interest: first, measure and then
compare the cross sections which will give a clue to the respective roles of
macroscopic and structure effects: charge asymmetry of the entrance channel and
closed shells effects. Secondly, produce ER's in different levels, especially
isomeric states.

$\it{-ii- New~isotopes~with~^{207}Pb:}$ Produce isotopes lighter by 1 neutron
than the ones known for several elements: $^{260}$Bh$_{107}$,
$^{265}$Mt$_{109}$, $^{271}$111 and $^{276}$112. For the last two cases, a
chain of 3 $\alpha$'s will be observed before reaching a known isotope, i.e. 3
new isotopes will be observed in one event.

$\it{-iii- New~elements:}$ Above Z=109 the counting rates are very low. For
$^{277}$112, measurements were made at one beam energy only, not necessarily at
the maximum of the excitation function. The broader energy coverage provided by
inverse kinematics may reveal a larger cross section. Moreover, the 1n channel
cross section decreasing quickly with Z, it may be that the 0n cross section
becomes larger even through the fusion probability drops very much at the
corresponding incident energy. Added to the uncertainty on the compound nucleus
mass defect, the exploration of a broad range of incident energies
corresponding to 0n+1n channels leads to lengthy measurements at several
incident energy in direct kinematics: 6 beam energies were tried for Se+Pb at
GSI in the search of element 116~\cite{ref1} whereas an excitation energy range
of 12 MeV or more can be covered with a single incident energy in inverse
kinematics.

\subsection{Use of $^{76}$Ge$_{32}$ beam}

The main interest of using $^{76}$Ge beam is the {\bf direct} cold fusion
production of the element $^{273}$110 using a $^{198}$Pt target~\cite{ref9}.
Indeed, this isotope has been observed in 3 events only through the decay of
element $^{277}$112 and possibly 3 events by direct production in {\bf hot}
fusion reaction ($^{34}$S+$^{244}$Pu=$^{273}$110+5n). A very large range of
$\alpha$-decay energies, from 9.18 to 11.17 MeV with lifetimes from 0.11 to 170
msec have been observed. What we propose is to make a precise
$\alpha$-spectroscopy of this isotope since the population of different levels
(especially isomeric states) may be due to overreaching the deformed shell
N=162. This would require a "large" number of events.

A more general interest for the $^{76}$Ge beam is provided by the accessible
region of production. Indeed, it makes possible, using targets ranging from
$^{192}$Os$_{76}$ to $^{205}$Tl$_{81}$, to produce elements 108 to 113 heavier
by two neutrons (except for element 111) than the isotopes produced using Pb/Bi
targets with beams ranging from $^{48}$Ca to $^{70}$Zn~\cite{ref1}. Clearly,
comparing with Pb/Bi based reactions, the entrance channel is more symmetric
and there are no closed shells, the cross sections will be then reduced. But on
the other hand, the accessible compound nuclei are richer by 2 neutrons (except
for element 111) and this will be in favour of the cross sections.

An estimation of the reduction factors (see Figure 3) can be made using the
compound nucleus $^{220}$Th produced via different entrance channels having
similar Bass barriers~\cite{ref4, ref10, ref11, ref12}: Ar$_{18}$+Hf$_{72}$ and
Zn$_{30}$+Nd$_{60}$ systems have no closed shells and the more symmetric system
has a cross section smaller by a factor 14. Assuming a steady variation with
the charge asymmetry, the expected cross sections for Ca$_{20}$+Yt$_{70}$ and
Zr$_{40}$+Sn$_{50}$ are estimated. The actual cross sections are much higher,
by a factor 6 and 23 respectively, due to the role of the closed
shells~\cite{ref13}. These factors may now be used to estimate the reduction
factor which applies when the same CN is formed on $^{76}$Ge based reactions in
comparison with Pb/Bi based reactions. The loss of shell effects is estimated
to reduce the cross section by one order of magnitude (represented by a solid
arrow in Figure 3). At Z=106, the decrease of asymmetry reduces the cross
section by a factor about 12 (represented by a dashed arrow in Figure 3)
whereas this asymmetry disappears for $^{76}$Ge+$^{208}$Pb reaction producing
element 114.

An estimation of the gain in the cross section due to the neutron richness can
be made also considering the Th isotopes. Detailed measurements show that 2
more neutrons in the CN increase the residue cross section by a factor
9~\cite{ref12}. As seen in Figure 3, the gain for Z=110 using $^{64}$Ni beam
instead of $^{62}$Ni is in the order of 5. In our estimation for $^{76}$Ge
beam, we consider an average factor of 7 (represented by a solid arrow on
Figure 3). Then, the estimated cross sections are represented by solid crosses.
They are smaller than the Pb/Bi based cross sections.

\subsection{Study of the structure of the Super Heavy Elements}

A new experimental program is proposed at GANIL~\cite{ref14} on the study of
transfermium isotopes by the use of spectroscopic information in order to have
a better understanding of their shell structure. This program is going to start
with the $\alpha$, $\gamma$ and electron spectroscopy of $^{251}$Md and
$^{251}$Fm. $^{251}$Md is populated by the $\alpha$-decay of $^{255}$Lr
produced in the 2 neutrons channel of the $^{48}$Ca+$^{209}$Bi fusion
evaporation reaction. The $^{251}$Md should decay with less than 10\% by
$\alpha$-decay on $^{247}$Es and with more than 90\% by EC on $^{251}$Fm. Since
the ground state parity and spin of the first level of $^{251}$Fm are known,
ground state of $^{251}$Fm and possibly of $^{255}$Lr could be determined. The
experimental setup will be adapted with a tunnel of Si detectors to detect
conversion-electrons and a set of Ge clover detectors from Exogam will be added
in a close geometry.

\section*{Acknowledgements}
The authors would like to acknowledge all the technicians and engineers from
mechanics, detector and electronic groups who made possible this experimental
program, especially at LPC Caen, GANIL and Saclay. We are also indebted to the
target laboratories of LNS Catania, IFU Sao Paulo, GSI Darmstadt and IPN Orsay
for the excellent quality of their products. The users support group (Service
des aires) and the accelerator staff of GANIL should be mentioned for their
continuous support and efficient performances at all stages of this work.
Finally, a great help was provided by the data acquisition group GIP (GANIL)
for the electronics and by the CHARISSA collaboration for lending us their VXI
ADC's.

\clearpage

\begin{table}
\caption{%
Summary of the different possible reactions using $^{207,208}$Pb beams on
various targets. Z$_{CN}$ and A$_{CN}$ are respectively the atomic number and
mass of the compound nucleus, E$^{*}_{bass}$ being the excitation energy at the
Bass barrier.} \label{table1}
\begin{center}
\begin{tabular}{c|ccc|ccc}
 & & $^{208}$Pb beam & & & $^{207}$Pb beam & \\
\hline
Z$_{CN}$ & Target & A$_{CN}$ & E$^{*}_{bass}$(MeV) & Target & A$_{CN}$ & E$^{*}_{bass}$(MeV)\\
\hline
104 & $^{50}$Ti & 258 & 23 & $^{50}$Ti & 257 & 23\\
105 &  $^{\it 51}$\it V$^{a}$ & \it 259 & \it 25 & $^{\it 51}$\it V & \it 258 & \it 25\\
106 & $^{54}$Cr & 262 & 22 & $^{54}$Cr & 262 & 22\\
107 & $^{\it 55}$\it Mn$^{a}$ & \it 263 & \it 24 & $^{\it 55}$\it Mn$^{b}$ & \it 262 & \it 24\\
108 & $^{58}$Fe & 266 & 21 & $^{58}$Fe & 265 & 20\\
109 & $^{\it 59}$\it Co$^{a}$ & \it 267 & \it 23 & $^{\it 59}$\it Co$^{b}$ & \it 266 & \it 22\\
110 & $^{64}$Ni & 272 & 17 & $^{64}$Ni & 271 & 16\\
111 & $^{\it 65}$\it Cu$^{a}$ & \it 273 & \it 19 & $^{\it 65}$\it Cu$^{b}$ & \it 272 & \it 18\\
112 & $^{70}$Zn & 278 & 12 & $^{\it 70}$\it Zn$^{b}$ & \it 277 & \it 12\\
113 & $^{\it 71}$\it Ga &\it  279 & \it 13 & $^{\it 71}$\it Ga & \it 278 & \it 13\\
114 & $^{\it 76}$\it Ge &\it  284 & \it 9 & $^{\it 76}$\it Ge & \it 283 & \it 9\\
115 & $^{\it 75}$\it As &\it  283 & \it 13 & $^{\it 75}$\it As & \it 282 & \it 13\\
116 & $^{\it 82}$\it Se &\it  290 & \it 3 & $^{\it 82}$\it Se & \it 289 & \it 2\\
117 & $^{\it 81}$\it Br &\it  289 & \it 7 & $^{\it 81}$\it Br & \it 288 & \it 6\\
\hline
\end{tabular}
\end{center}
{\it new reactions}, $^{a}$ new path,
$^{b}$ production of new isotope\\

\end{table}


\begin{figure}
\includegraphics[height=15cm,keepaspectratio,angle=-90]{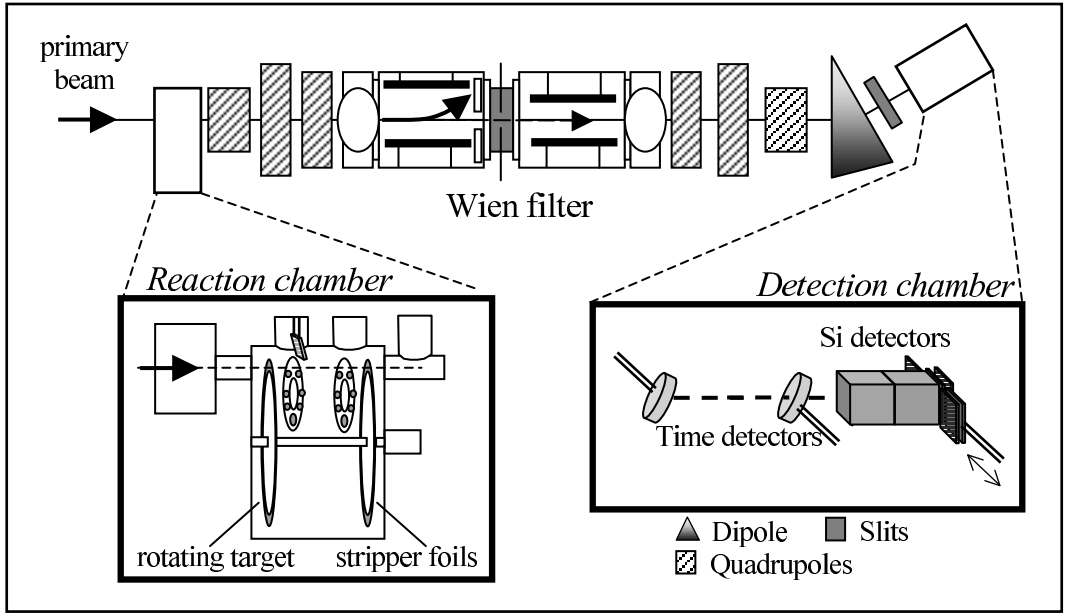}
\caption[]{%
Schematic drawing of the experimental setup including, from left to right, the
reaction chamber with the rotating target and stripper foils, the Q-poles and
the LISE3 Wien filter between them, the last dipole and the detection chamber
with time and Si detectors. } \label{fig1}
\end{figure}%

\begin{figure}
\includegraphics[height=15cm,keepaspectratio,angle=-90]{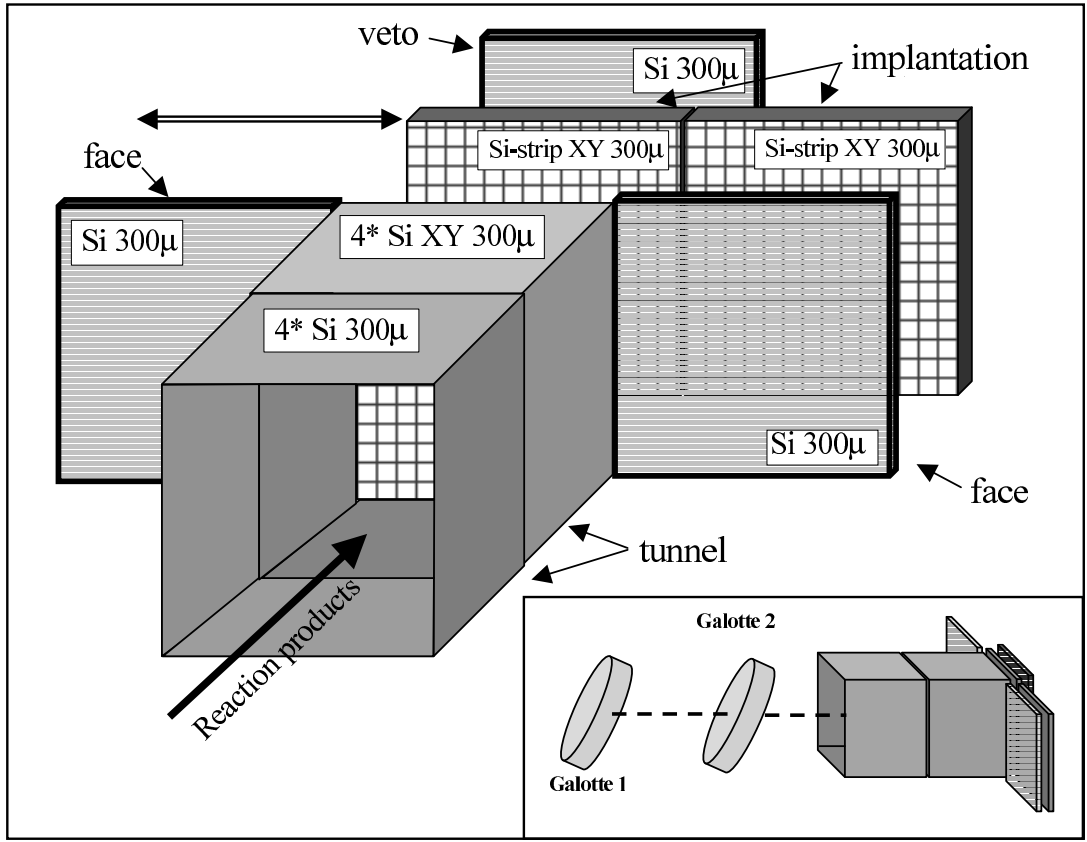}
\caption[]{%
Schematic view of the detection setup implanted at the end of the LISE3
spectrometer. It is composed by 2 time detectors (Galotte), 2 movable
implantation Si detectors, a set of 8 Si detectors for the tunnel, 1 veto Si
detector behind the 'in beam' implantation and 2 Si detectors (face) to recover
the products ($\alpha$ or fission fragment) escaping the 'out of beam'
implantation detector. } \label{fig2}
\end{figure}%

\begin{figure}
\includegraphics[height=15cm,keepaspectratio,angle=-90]{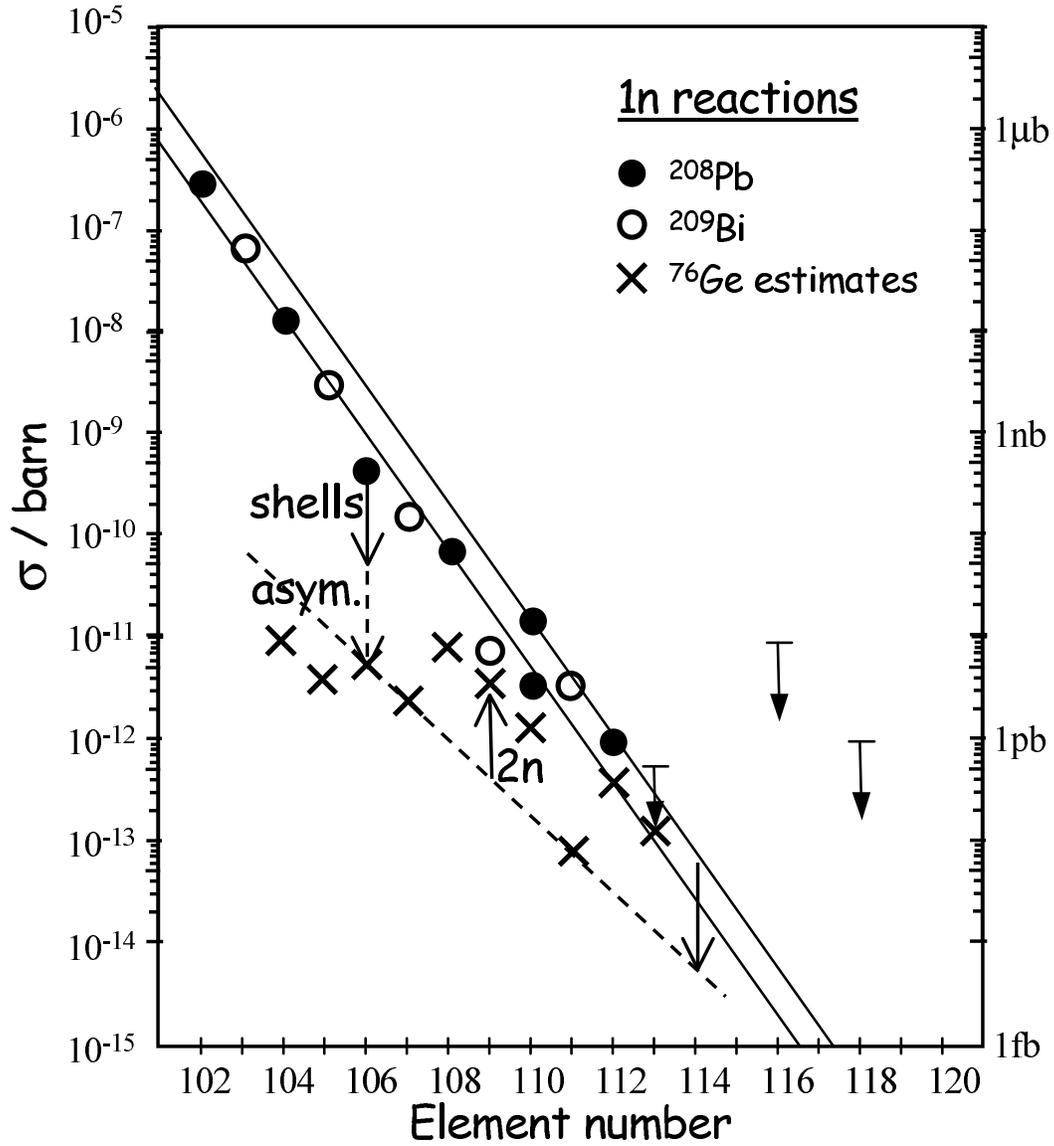}
\caption[]{%
One neutron channel cross sections measured for $^{208}$Pb and $^{209}$Bi
targets in direct kinematics~\cite{ref6}. The crosses represent our estimations
using a $^{76}$Ge beam (see text sect. 4.2. for details). } \label{fig3}
\end{figure}%

\end{document}